\documentclass[lettersize,journal]{IEEEtran}
\usepackage{multirow}
\usepackage{amsmath,amsfonts}
\usepackage{algorithm}
\usepackage{graphicx}
\usepackage{textcomp}
\usepackage{xcolor}
\usepackage{mathrsfs}
\usepackage{booktabs}
\usepackage{url}
\usepackage{algpseudocode}
\algrenewcommand\algorithmicrequire{\textbf{Input:}}
\algrenewcommand\algorithmicensure{\textbf{Output:}}
\usepackage{amsthm}
\newtheorem{theorem}{Theorem}
\newtheorem{lemma}{Lemma}

\newtheorem{definition}{Definition}

\usepackage{flushend}
\def\BibTeX{{\rm B\kern-.05em{\sc i\kern-.025em b}\kern-.08em
    T\kern-.1667em\lower.7ex\hbox{E}\kern-.125emX}}
\renewcommand{\thesection}{\arabic{section}}  
\newcounter{OPequ}
\newenvironment{CEquation}
  {\stepcounter{OPequ}%
    \addtocounter{equation}{-1}%
    \renewcommand\theequation{OP\arabic{OPequ}}\equation}
  {\endequation}
\usepackage{mathtools}
  
\usepackage{subfig}
\usepackage[normalem]{ulem}
\usepackage{scalerel}
\usepackage{tikz}
\usetikzlibrary{svg.path}
\definecolor{orcidlogocol}{HTML}{A6CE39}
\tikzset{
  orcidlogo/.pic={
    \fill[orcidlogocol] svg{M256,128c0,70.7-57.3,128-128,128C57.3,256,0,198.7,0,128C0,57.3,57.3,0,128,0C198.7,0,256,57.3,256,128z};
    \fill[white] svg{M86.3,186.2H70.9V79.1h15.4v48.4V186.2z}
                 svg{M108.9,79.1h41.6c39.6,0,57,28.3,57,53.6c0,27.5-21.5,53.6-56.8,53.6h-41.8V79.1z M124.3,172.4h24.5c34.9,0,42.9-26.5,42.9-39.7c0-21.5-13.7-39.7-43.7-39.7h-23.7V172.4z}
                 svg{M88.7,56.8c0,5.5-4.5,10.1-10.1,10.1c-5.6,0-10.1-4.6-10.1-10.1c0-5.6,4.5-10.1,10.1-10.1C84.2,46.7,88.7,51.3,88.7,56.8z};
  }
}

\newcommand\orcidicon[1]{\href{https://orcid.org/#1}{\mbox{\scalerel*{
\begin{tikzpicture}[yscale=-1,transform shape]
\pic{orcidlogo};
\end{tikzpicture}
}{|}}}}

\usepackage{hyperref}
\begin{document}

\title{Dimension Reduction via Random Projection for Privacy in Multi-Agent Systems
{\footnotesize \textsuperscript{}}

}

\author{Puspanjali Ghoshal \orcidicon{0000-0002-2324-7001} , ~\IEEEmembership{Member, IEEE}, Ashok Singh Sairam \orcidicon{0000-0001-9527-6496}\,, ~\IEEEmembership{Senior Member, IEEE}
        % <-this % stops a space
\thanks{Puspanjali Ghoshal is a research scholar in the Department of Mathematics, Indian Institute of Technology, Guwahati, Assam, India.(email: g.puspanjali@iitg.ac.in)}% <-this % stops a space
\thanks{Ashok Singh Sairam is a professor in the Department of Mathematics, Indian Institute of Technology, Guwahati, Assam, India.(email: ashok@iitg.ac.in)}}

\maketitle

\begin{abstract}
  In a Multi-Agent System (MAS), autonomous agents observe various aspects of the environment and transmit this information to a central entity responsible for aggregating the data and deducing system parameters. To improve overall efficiency, agents may append certain private parameters to their observations. For example, in a crowd-sourced traffic monitoring system, commuters might share not only their current speed, but also sensitive information such as their location to enable more accurate route prediction. However, sharing such data can allow the central entity or a potential adversary to infer private details about the user, such as their daily routines. To mitigate these privacy risks, the agents sanitize the data before transmission. This sanitization inevitably results in a loss of utility. In this work, we formulate the problem as a utility-privacy trade-off and propose a novel compression-based approach leveraging the notion of robust concepts to sanitize the shared data. We further derive a bound on the norm of the compression matrix required to ensure maximal privacy while satisfying predefined utility constraints.
\end{abstract}


\maketitle

\begin{sloppypar}
\section{Introduction}
A Multi-agent system (MAS)\cite{MAS_Survey} consists of multiple autonomous agents (for example, sensors) that sense and report data on system parameters. This data is sent to a fusion center, where it is aggregated to accurately assess the parameters. MASs have found widespread application in multiple domains such as weather forecasting\cite{weather}, surveillance and healthcare\cite{health}. The data communicated to the fusion center includes some private parameters of the agents which augment interpretation and analysis of the monitored system. The aim of this work is to achieve \textit{inference privacy}\cite{Identity_inference}, a relatively new privacy concept. Cryptographic privacy provides protection of data from a malicious third party. Inference privacy, on the other hand, refers to the ability of protecting sensitive information from being inferred by attackers from shared private data.

Consider that a cosmetic company that wishes to conduct a survey about its products. They create a questionnaire and offer a discount voucher for every successful completion of the survey. The poll has questions related to the products, along with age, income level, and email id for the voucher. The people who complete the questionnaire are the agents, and the company is the fusion center. The company aggregates the data to analyze the popularity of the product. Cryptographic privacy in this context would imply that only authorized personnel would be able to access the responses. Inference privacy, on the other hand, requires that no one be able to create a profile of the user, such as inferring lifestyle or social status. 

Apart from MAS, inference privacy preservation is also crucial in other domains, like Location-based Services, where the user location trajectory can be exploited to infer secondary information like their job and place of residence. In \cite{R6}, authors ensure inference privacy preservation in the context of location trajectory based data sharing, ensuring privacy preservation by a two step optimization process. However, in the case of MAS, the involvement of multiple private parameters, including, but not restricted to, location data, necessitates a more careful and comprehensive design of the privacy mechanism. 

Ensuring inference privacy requires data obfuscation, which inevitably results in the loss of data utility. To quantify the levels of utility and privacy and formulate the trade-off problem, advanced statistical functions have been utilized in the literature \cite{ASUP} \cite{Desired_util_priv}. In this work, we propose to quantify utility and privacy using \textit{cosine similarity}, which has a computational complexity dependent linearly on the vector length \cite{cosSim}. The definition proposed in this paper, is different from earlier definitions using cosine similarity \cite{ANTS2023}. The earlier definitions quantified utility and privacy of the system as a whole using the sum and products of the cosine similarities. However, in a decentralized setup, the quantification of agent utility and privacy is of equal importance. We thus, have defined in this paper, the utility and privacy of each agent. To validate our definition, we demonstrate that all established results can be derived using it. Next, we formulated the problem of enforcing inference privacy as one of maximizing privacy while maintaining a given utility level. 

There are two main approaches for enforcing inference privacy: \textit{noise-addition based} and \textit{compression-based}. Compression-based methods involve the dimensionality reduction technique, one of which is matrix multiplication. \textit{Random projection} method is a dimensionality reduction method used to modify \textit{examples} of a \textit{Robust Concept}\cite{Random_Proj1}. Our problem shares a strong analogy with the deduction of a Robust Concept, which involves estimating the attributes of a phenomenon based on examples that are altered up to a specified degree. In a MAS, the system parameters estimated by the fusion center correspond to the attributes of the concept, while the agent observations correspond to the examples. In this work, we propose a novel matrix multiplication-based random projection for data sanitization to enforce inference privacy. We derive a matrix norm bound that satisfies the utility constraint. This bound is incorporated into the matrix generation phase of our proposed sanitization function.

Our contributions can be summarized as follows.
\begin{itemize}
    \item Formulation of the MAS system model as a problem of deducing a robust concept from examples.
    \item Deduction of a bound on the compression matrix norm for achieving maximum privacy under set utility levels.
    \item We propose a decentralized sanitization mechanism for our MAS privacy problem using a variant of \textit{Random projection}.
    \item Experimentally show the efficiency of our method over existing methods.
\end{itemize}

The rest of the paper is organised as follows. The literature survey followed by the system model is given in Section \ref{sec:survey} and Section \ref{Sys_model}, respectively. In Section \ref{sec:Methodology}, the sanitization mechanism is presented. We then have the experiments and results in Section \ref{sec:expt} followed by the future research direction in Section \ref{Future}.

\section{Related Work}\label{sec:survey}
Privacy requirements in a MAS have been extensively explored in literature \cite{NewSurvey1} \cite{NewSurvey2}. In this section, we review the methods for enforcing inference privacy, which can be broadly classified into two categories: noise-addition-based methods and compression-based methods. The process of enforcing inference privacy comes at the cost of reduced utility of the data. We further examine how the literature has addressed the resulting utility-privacy trade-off.
\subsection{Noise-addition based methods}
Noise-addition based methods perturb the data before transmission to mask personal-level information \cite{noise1}. The noise is randomly generated from a suitable probability distribution. Commonly used noise-addition mechanisms include the Laplacian method and the Gaussian method\cite{Dwork}\cite{noise2}\cite{noise3}. 
Although these methods provide a strong privacy guarantee under formal frameworks, such as differential privacy, they suffer from significant utility degradation. A high magnitude of noise is required to achieve higher privacy values. Consequently, the data undergo significant modification and thus the properties of the raw data are not retained in the sanitized data. In contrast, compression-based methods aim to reduce the amount of shared information without injecting randomness, thereby offering a more controlled utility-privacy trade-off.
\subsection{Compression-based methods}
Compression-based methods reduce the dimensionality of the data  prior to communication, thereby limiting the amount of information that can be inferred from each transmission. It tries to find a balance between privacy and utility either by retaining only the most significant features or by encapsulating all features within a reduced-dimensional representation. A classical example of such an approach is Principal Component Analysis (PCA) \cite{Compression5,New_PCA}, which linearly projects high-dimensional data onto a lower-dimensional subspace defined by the directions of greatest variance. PCA identifies and retains the most informative components, effectively reducing the number of features while preserving much of the original structure. However, a major limitation of PCA is its reliance on access to the entire dataset to compute the principal components. This makes it unsuitable for online or decentralized settings such as MAS, where agents transmit data one tuple at a time without the presence of a Trusted Third Party (TTP). Moreover, owing to the limited computational resources of the agents, training a decentralized PCA model is infeasible.

The other class of compression-based methods, the main focus of this paper, projects the data tuples to a lower-dimensional space, effectively encapsulating all relevant features in the resulting tuple. This is typically achieved by matrix multiplication, where a high-dimensional data vector is multiplied by a projection matrix to produce its compressed representation. 
Zhou et al.~\cite{Compression1} proposed multiplying the entire data matrix with a matrix whose entries are drawn from a Gaussian distribution. In a follow-up study\cite{Compression1'}, it was demonstrated that such compressed data retain enough information to deduce system parameters with accuracy comparable to the original dataset.

A common drawback of the compression-based methods discussed so far is that they assume access to a complete dataset, or at least a significant portion of it, before the compression can be applied. In contrast, agents in an MAS typically transmit individual tuples in real time, without access to the full dataset, making these methods practically infeasible. The problem is partially addressed by Basic Random Projection (BRP)~\cite{Random_Proj1}. It uses a fixed uniform random orthonormal matrix to project individual data tuples to a lower-dimensional subspace. This matrix effectively serves as a basis for the lower-dimensional space. The second advantage is that smaller dimensional messages are easier to transmit and use less resources of the system.

Although BRP eliminates the need for the entire dataset, it introduces computational overhead due to the requirement of an orthonormal matrix, which may not be feasible for agents with limited processing capabilities. Our proposed sanitization approach also uses a matrix for compression, however, the matrix is formed by independent entries from a bounded distribution. Eliminating the requirement of an orthonormal matrix reduces the load on the limited resources of the agents.

\subsection{Utility-Privacy tradeoff problem}
To ensure privacy, the data observed by the agents is sanitized before transmission either by adding noise or compression. This modification inevitably leads to a loss in utility, the quality or usefulness of the data for deriving information at the fusion center. As such, there exists an inherent trade-off between utility and privacy, which must be carefully balanced based on the system's objectives and the agents' resource constraints. 

A classical approach to addressing this trade-off is to formally define mathematical expressions for both utility and privacy, and pose the problem as one of constrained optimization \cite{wangOpti}. Wang et al. \cite{ASUP} introduced a framework based on the \textit{Cramér-Rao lower bound } (CRLB) and the \textit{Fisher Information Matrix (FIM)} to quantify utility and privacy. They proposed two flavors of the optimization problem: one over the space of all possible sanitization functions, and another over the noise covariance matrix in the case of noise-addition mechanisms. The authors proposed a noise-addition based sanitization mechanism, \textit{Arbitrarily strong privacy with perfect utility sanitization algorithm} (ASUP algorithm), that has been shown to attain privacy under the constraint of preserving perfect utility. However, a significant limitation of their approach is its high computational cost. The worst-case complexity of computing the CRLB is $O(n^3)$\cite{CRLB} which is computationally demanding for resource-constrained agents. 
 
Wang et al.~\cite{Desired_util_priv} define the problem as the identification of a sanitization mechanism that balances utility loss with privacy gain. They do this by assuming average information leakage as the privacy metric and use probabilistic mapping for sanitization. This approach requires significant computational ability. However, these studies focus on either optimization formulation or privacy enforcement, without integrating both into the problem definition. In this paper, we bridge the gap by defining the outcome of utility-privacy optimization directly into a compression-based privacy mechanism.

\section{System Model}\label{Sys_model}
In this section, we first present the system model and formally define data utility and privacy. We then formulate the problem as a utility–privacy trade-off. 

Consider an MAS of $N$ agents deployed in an observation area of $L \times L$. We assume, without loss of generality, that the area is divided into equal sized square grid cells of length $l$, each containing one agent. An agent $i$ gathers information about the system parameter tuple $\mathbf{x} \in \mathbb{R}^q$ and reports an observation vector $\mathbf{y_i} \in \mathbb{R}^n$ to the fusion center.  The parameter $q$ is the dimension of the system parameter that the fusion center will infer after aggregating all agent data of dimension $n$. Since agents may observe part of the whole, $q \ne n$ is possible. The fusion center receives the data from all agents to accurately conclude about $\mathbf{x}$. 
 
\begin{figure}[h]
    \centering
    \includegraphics[width=0.75\linewidth]{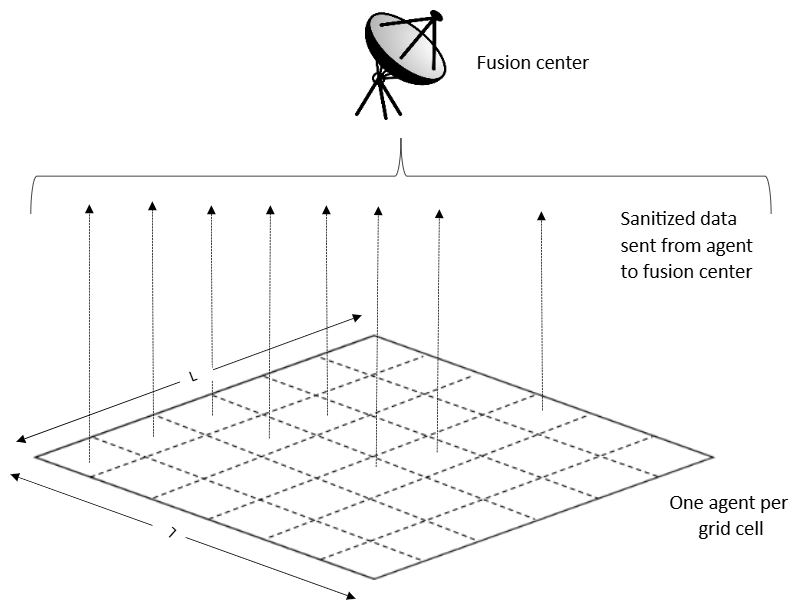}
    \caption{MAS Architecture}
    \label{MAS_Arch}
\end{figure}

The $n$-tuple observation vector, $\mathbf{y_i}$, contains the observation data as well as private parameters of the agent. Sending this data directly to the fusion center poses privacy risks. To obfuscate the data and reduce susceptibility to inference attacks, the agents sanitize the data vector $\mathbf{y_i}$ using a decentralized privacy algorithm.
The architecture of a MAS is shown in Figure \ref{MAS_Arch}.

The data tuple structure is assumed to be uniform across all the agents, that is, the position of the private parameters within the data tuple is the same for all agents. Each agent executes privacy mechanism $T$ in a decentralized manner. The data communicated by the agent to the fusion center is represented in Equation \ref{eqn:T(y)},
\begin{equation}\label{eqn:T(y)}
    T(\mathbf{y_i}) \in \mathbb{R}^m,
\end{equation}
where $m \leq n$, indicating that the dimension of the sanitized data may be smaller than the dimension of the raw data, depending on the sanitization function used. 

\subsection{Quantifying utility and privacy:}
To protect data from inference privacy breaches, sanitization techniques, such as noise addition or dimensionality reduction, are needed. In doing so, raw data are often distorted, thereby degrading the utility of the data. Quantifying this loss is essential for evaluating the effectiveness of a privacy mechanism. It enables system designers to find a trade-off between utility and privacy. To this end, we use the cosine similarity between the raw and sanitized data vectors as a unified measure of both utility and privacy. Computation of cosine similarity requires vectors of equal length. If $T$ reduces dimension ($m < n$), the cosine similarity can be computed by embedding $T(\mathbf{y_i})$ back into $\mathbb{R}^n$ (via zero padding or reconstruction).

Cosine similarity has previously been used to measure the plagiarism between documents \cite{PlagCosSim} %\cite{Cos1}
and for face verification\cite{Cos2}. Its successful application in these diverse domains justifies its use in our context to quantify the loss in data utility introduced by sanitization. The similarity measure indicates how closely the sanitized data aligns with the original in terms of direction, with a higher cosine similarity implying greater similarity between the two vectors. The similarity between vectors $\mathbf{x}$ and $\mathbf{y}$, denoted as $cos(\mathbf{x},\mathbf{y})$, is computed as given in Equation \ref{eq:cos},
\begin{equation}
    \label{eq:cos}
    cos(\mathbf{x},\mathbf{y})=\frac{\mathbf{x} \cdot \mathbf{y}}{||\mathbf{x}|| \times ||\mathbf{y}||}.
\end{equation}
Cosine similarity is invariant to scaling. This property ensures that, even if the sanitization process alters the magnitude of the data, the underlying directional information remains a reliable measure of utility preservation. To restrict the value of Cosine similarity between 0 and 1, as shown in Equation~\ref{eqn:Constraint} we make the following assumptions. In noise addition-based mechanisms, the noise added is assumed to be independent of the underlying dataset. In compression-based methods that employ a matrix (say, $A$) is generally used for dimension reduction \cite{Random_Proj1}, \cite{Compression1}, \cite{Compression5}, the matrix is assumed to be orthonormal.  The matrix $A$ acts as a basis for the lower-dimensional space. In case the matrix is not positive semidefinite, we can take $A^TA$ for compression and drop the extra rows not originally present in the compression matrix. 
\begin{equation}\label{eqn:Constraint}
    0 \leq cos(\mathbf{y_i}, T(\mathbf{y_i})) \leq 1.
\end{equation} 

Equation~\ref{eqn:Constraint} follows directly from the definition of cosine similarity, the range that generally lies between -1 and 1. However, since both $\mathbf{y_i}$ and $T(\mathbf{y_i})$ share the same quadrant due to the projection mechanism used, the cosine similarity is restricted to the interval [0,1]. 

Using the above results, we have the following definitions of utility and privacy:
\begin{definition}
    Utility of agent data: The utility of the data belonging to the agent $i$ after application of the sanitization function $T$ is given as $u_i(T)=cos(\mathbf{y_i}, T(\mathbf{y_i})).$
\end{definition}
 As the similarity between the tuples increases, the utility of the data increases. We say that the sanitized data provides perfect utility when the value of utility is 1.
\begin{definition}
    Agent privacy: The privacy achieved by agent \textit{i} due to the application of the sanitization function $T$ is given as $p_i(T)=1-cos(\mathbf{y_i}, T(\mathbf{y_i}))=1-u_i(T).$
\end{definition}
As the utility of the data increases, the level of privacy decreases. From this definition, we observe that privacy decreases as the similarity between tuples increases.  Since $cos(\mathbf{y_i}, T(\mathbf{y_i})) \in [0,1]$, we subtract the similarity value from 1 to determine the level of privacy achieved. 

We directly quantify $p_i(T)=1-u_i(T)$ without the introduction of any other scaling parameter so that our formulation remains interpretable, and maps directly to the geometric notion of cosine similarity. The privacy of an agent can also be alternatively defined as $p_i(T)=\alpha(1-u_i(T)), \text{ with } \alpha >0$, in cases where there is an asymmetry in the tradeoff. In many situations, modifying or disclosing certain aspects of the data can have a disproportionate effect on privacy compared to the resulting utility. That is, privacy and utility may respond to data changes at different rates, and their relationship may not be linearly inverse. In certain healthcare, finance and social media scenarios, some parameters are far more privacy sensitive than others; even if those features significantly enhance utility, their disclosure may be unacceptable. In such cases, the parameter $\alpha$ provides flexibility and captures this asymmetry.
However, the value of $\alpha$ is domain and design specific, requiring additional justification or empirical evidence. Without such evidence, introducing a scaling parameter could obscure the direct relationship between privacy and utility.
\subsection{Comparative analysis of utility and privacy}
To validate our definitions of utility and privacy, we compare them with those proposed in the literature. In particular, we focus on the definition proposed by Wang et al. \cite{ASUP}, which make use of CRLB and FIM. 
Continuing with our earlier definitions of $\mathbf{x}$ as the system parameter and $\mathbf{y}$ as the agent's observation, we now examine the formulation proposed in Wang et al. \cite{ASUP}. In their framework, the CRLB for estimating $\mathbf{x}$ is defined as
\[ 
 P_\mathbf{x} := (J_\mathbf{x} + J_0)^{-1}
 \]
where $J_\mathbf{x}$ and $J_0$ are Fisher information matrices. Taking $T$ as the sanitization function, let $\tilde{\mathbf{y}} = T(\mathbf{y})$ be the sanitized version of $\mathbf{y}$. The corresponding Fisher information is denoted as $J_T$ and the updated CRLB is given by
\[
\widetilde{P}_\mathbf{x} := (J_T + J_0)^{-1}.
\]
In Theorem~\ref{thm:cos_vs_crlb}, we show that our proposed utility and privacy measures are bounded above by CRLB-based measures. The sanitized data are expressed as $ \tilde{\mathbf{y}} = \mathbf{y} + \delta\mathbf{y} $ with $\delta\mathbf{y}$ denoting modification by compression or noise addition. 

\begin{theorem}\label{thm:cos_vs_crlb}
Let $\mathcal{P}_{\mathrm{CRLB}} \text{ and } \mathcal{U}_{\mathrm{CRLB}}$  be the privacy and utility values, respectively, computed using the CRLB framework. Similarly, let $\mathcal{P}_{\mathrm{cos}} \text{ and } \mathcal{U}_{\mathrm{cos}}$ denote the corresponding privacy and utility values computed using cosine similarity. Then under the assumption $\cos(\mathbf{y}, \tilde{\mathbf{y}}) \to 1$, there exist constants $ C_1>0 \, \text{ and } C_2<0$ such that:
\begin{equation}\label{Cos_CRLB}
\left\{
\begin{array}{lll}
\displaystyle \mathcal{P}_{\mathrm{cos}} \leq C_1 \cdot \mathcal{P}_{\mathrm{CRLB}} + o(\mathcal{P}_{\mathrm{CRLB}}), \\[6pt]%1 - 
\displaystyle \mathcal{U}_{\mathrm{cos}} \leq C_2 \cdot \mathcal{U}_{\mathrm{CRLB}} +o(1).
\end{array}
\right.
\end{equation}
The constants depend on the observation model, sanitization mechanism, and the availability of prior information.
\end{theorem}
\begin{proof}
We have \( \delta \mathbf{y} := \tilde{\mathbf{y}} - \mathbf{y} \), and assume \( \| \delta \mathbf{y} \| \ll \|\mathbf{y}\| \). By Taylor series expansion, we have:
\begin{equation*}
    \cos(\mathbf{y}, \tilde{\mathbf{y}}) = \frac{\mathbf{y}^T (\mathbf{y} + \delta \mathbf{y})}{\|\mathbf{y}\| \cdot \|\mathbf{y} + \delta \mathbf{y}\|} \approx 1 - \frac{1}{2} \cdot \frac{\|\delta \mathbf{y}\|^2}{\|\mathbf{y}\|^2}.
\end{equation*}
Therefore:
\begin{equation*}
    1 - \cos(\mathbf{y}, \tilde{\mathbf{y}}) \approx \frac{1}{2} \cdot \frac{\|\delta \mathbf{y}\|^2}{\|\mathbf{y}\|^2}.
\end{equation*}
Data sanitization reduces Fisher information as follows:
\begin{equation*}
    J_T = J_\mathbf{x} - \Delta, \quad \text{with } \|\Delta\| \sim O(\|\delta \mathbf{y}\|^2).
\end{equation*}
This gives:
\begin{equation*}
    \begin{split}
        \widetilde{P}_\mathbf{x} & = (J_T + J_0)^{-1} \\
        &= (J_\mathbf{x} + J_0 - \Delta)^{-1} \\ &\approx P_\mathbf{x} + P_\mathbf{x} \Delta P_\mathbf{x} + o(\|\delta \mathbf{y}\|^2)
    \end{split}
\end{equation*}
Considering $U$ and $G$ to be the matrices specifying the public and private parameters, then:
\begin{equation*}
    \begin{split}
        &\mathcal{P}_{\mathrm{CRLB}} \approx \frac{\operatorname{Tr}(G P_\mathbf{x} \Delta P_\mathbf{x} G^T)}{\operatorname{Tr}(G P_\mathbf{x} G^T)} \sim O(\|\delta \mathbf{y}\|^2)\\
\Rightarrow \; &
1 - \cos(\mathbf{y}, \tilde{\mathbf{y}}) \leq C_1 \cdot \mathcal{P}_{\mathrm{CRLB}} + o(\mathcal{P}_{\mathrm{CRLB}})\\
\Rightarrow \; &
\mathcal{P}_{\mathrm{cos}} \leq C_1 \cdot \mathcal{P}_{\mathrm{CRLB}} + o(\mathcal{P}_{\mathrm{CRLB}})
    \end{split}
\end{equation*}
Similarly, for utility:
\begin{equation*}
    \mathcal{U}_{\mathrm{CRLB}} \approx -\frac{\operatorname{Tr}(U P_\mathbf{x} \Delta P_\mathbf{x} U^T)}{\operatorname{Tr}(U P_\mathbf{x} U^T)}
\end{equation*}
\text{Thus, } 
\begin{equation*}
    \begin{split}
    &|\mathcal{U}_{\mathrm{CRLB}}| = O(\|\delta \mathbf{y}\|^2), \; \text{and } \mathcal{U}_{\mathrm{CRLB}} \leq 0, \\
\Rightarrow \;&
\mathcal{U}_{\mathrm{cos}} \leq C_2 \cdot \mathcal{U}_{\mathrm{CRLB}} +o(1).
    \end{split}
\end{equation*}
\end{proof}

CRLB-based methods require access to model-specific Fisher Information Matrices and can be computationally intensive. In contrast, cosine similarity is straightforward to compute. The theorem suggests that our measures offer a practical and interpretable approximation to the CRLB-based metrics.

\subsection{Problem Definition} 
We now formally state the utility-privacy trade-off problem. In the literature, this trade-off is known to be application-specific \cite{Survey_MAS_Privacy}. That is, the permissible trade-off depends on the goal of the system. In our setting, the aim is to maximize agent-wise privacy while ensuring that predefined utility requirements are met. Mathematically, we want a matrix multiplication-based sanitization function $T \in \mathscr{T}$ such that
\begin{CEquation}\label{OP1}
        \begin{aligned}
        &\max_{T \in \mathscr{T}} && p_i(T),\\
        &\textrm{s.t.} && u_i(T)\geq \epsilon_i. %u_i(T)=1
        \end{aligned}
\end{CEquation}
The parameter $\epsilon_i>0$ is the utility requirement of the agent $i$ and $\mathscr{T}$ is the set of all privacy mechanisms. Rewriting the constraint using Equation \ref{eqn:Constraint}, we have: 
\begin{equation}\label{eqn:Constraintnew}
    \epsilon_i \leq cos(\mathbf{y_i},T(\mathbf{y_i})) \leq 1.
\end{equation}

\section{Inference privacy preservation using random projection}\label{sec:Methodology}
In this section, we describe the proposed sanitization mechanism based on random projection. The approach employs a compression matrix to achieve the desired privacy-utility trade-off. We begin by deriving a bound on the norm of the compression matrix necessary to meet the specified trade-off. This bound is then incorporated into the matrix generation step of our mechanism. Next, we introduce the relevant concepts and definitions. Finally, we present the neuronal random projection technique.

\subsection{Deriving Norm Bounds for the Compression Matrix}
We use the inequality given in Equation \ref{eqn:Constraintnew} to derive a relationship between the raw data $\mathbf{y_i}$ and its sanitized counterpart $T(\mathbf{y_i})$. These relations subsequently help to find the bounds on the norm of the compression matrix necessary to achieve the desired utility range. %perfect utility. 
The ideal perfect utility scenario is $cos(\mathbf{y_i} T(\mathbf{y_i}))=1$. This implies that $T(\mathbf{y_i})$ and $\mathbf{y_i}$ will be collinear, that is, one will be a positive scalar multiple of the other. Let that positive scalar be $t$. Thus we have, 
\begin{equation*}
    \exists t>0 \textrm{ s.t. } T(\mathbf{y_i})=t\mathbf{y_i}.
\end{equation*}
Now, the value of $t$ needs to be such that the distance between the sanitized data $T(\mathbf{y_i})$ and $\mathbf{y_i}$ does not exceed the length of a grid cell. We thus have
\begin{equation*}
        \begin{aligned}
            |t-1|||\mathbf{y_i}|| \leq l. \\
        \end{aligned}
\end{equation*}
Given that all tuples are bounded, for agent $i$, let $||\mathbf{y}||$ denote the maximum possible length of the tuple. We then obtain
\begin{equation}
        \begin{aligned}
            t \leq \frac{l}{||\mathbf{y}||}+1.
        \end{aligned}
\end{equation}
In the following theorem, we deduce the bounds on the norm of the compression matrix for the given range of utility.

\begin{theorem}{Given $T$ is the sanitization mechanism, the compression matrix $A_i$ used by agent $i$ should be such that $0 < ||A_i||_F \leq \min\{t, 2\epsilon_i +\delta \}$, where $\delta=O(l/||\mathbf{y}||)$.}\label{Lemma:Norm_bound}
\end{theorem}
\begin{proof}
    The sanitization function can be defined as $T(\mathbf{y_i})=A_i^T\mathbf{y_i}$, so Equation \ref{eqn:Constraintnew} can be reformulated as:
    \begin{equation}\label{eqn5}
        \begin{aligned}
        \epsilon_i \leq cos(\mathbf{y_i}, A_i^T\mathbf{y_i}) \leq 1.
        \end{aligned}
    \end{equation}
    We solve it using the properties of cosine similarity and the properties of norm. The ideal case of Equation \ref{eqn5} is $cos(\mathbf{y_i}, A_i^T\mathbf{y_i})$=$1$. This implies that $A_i^T\mathbf{y_i}$ and $\mathbf{y_i}$ are collinear. From the definition of Frobenius norm and the assumption that the compression matrix is positive semidefinite, we have the following bound on the norm of the matrix $A_i$:
     \begin{equation*}
        \begin{aligned}
            k_{max} \leq ||A_i||_F \leq \sqrt{n}k_{max},
        \end{aligned}
    \end{equation*} where $k_{max}>0$ is the largest eigen value of a non-zero positive semidefinite matrix.
    Since $t$ is the maximum modification that any vector can undergo by sanitization, we have the following bound on $||A_i||_F$:  
    \begin{equation}\label{bound1}
        \begin{aligned}
            0 < k_{max} \leq ||A_i||_F \leq \sqrt{n}k_{max} \leq t.
        \end{aligned}
    \end{equation}
Again, from $cos(\mathbf{y_i}, A_i^T\mathbf{y_i}) \geq \epsilon_i$ and $||A_i^T\mathbf{y_i} - \mathbf{y_i}||\leq t$ and assuming $||\mathbf{y_i}||=\alpha$, we have 
\begin{equation*}
    ||A_i^T\mathbf{y_i}||^2+\alpha^2-2\epsilon_i\alpha||A_i^T\mathbf{y_i}||\leq t^2.
\end{equation*}
We solve the above quadratic equation in $||A_i^T\mathbf{y}_i||$ to get an upper bound on $||A_i^T\mathbf{y}_i||$. From the inequality $||A_i^T\mathbf{y_i}|| \leq ||A_i||_F||\mathbf{y_i}||$ we can say that $||A_i||_F||\mathbf{y_i}||$ acts as the least upper bound of $||A_i^T\mathbf{y}_i||$. Thus we get 
\begin{equation*}
    ||A_i||_F\leq \epsilon_i +\sqrt{\epsilon_i^2-(1-t^2/\alpha^2)}.
\end{equation*}
Substituting the value of $t$, we have 
\begin{equation}\label{bound2}
    ||A_i||_F\leq 2\epsilon_i+\delta, \text{ where $\delta$= $O(l/||\mathbf{y}||)$}.
\end{equation}
Combining the bounds obtained in Equation \ref{bound1} and Equation \ref{bound2}, we have the final bound as
\begin{equation}
    0 < ||A_i||_F \leq \min\{t, 2\epsilon_i +\delta \}.
\end{equation}
\end{proof}
Having established bounds on the norm of the compression matrix, we now draw an analogy between our problem and the task of deducing a robust concept. In prior work, this notion has been explored in the context of dimensionality reduction through random projections \cite{Random_Proj1}. In our framework, we adapt this approach to enforce privacy by incorporating the derived norm bounds into the design of the compression matrix.

\subsection{Robust Concepts}\label{Prob_form} 
We formally define the notion of a concept,  followed by the definition of a robust concept, and draw parallels with the MAS problem. 
\begin{definition}(Concept)
    A concept is an idea of a physical or abstract phenomenon. It is formed by combining examples that contribute to understanding the attributes of the concept. The examples are combined following some preset rules. 
\end{definition}
For example, if $\mathbf{y_1}, \mathbf{y_2}, \ldots, \mathbf{y_N}$ are positive examples for a concept $\mathbf{x}$, then
\begin{equation*}
    \mathbf{x} = F(\mathbf{y_1}, \mathbf{y_2}, \ldots, \mathbf{y_N}),
\end{equation*}where $F$ is the function used to deduce the concept. In a MAS setup, the system parameter which the fusion center wishes to deduce is analogous to a \textit{concept}. In order to deduce the concept, the fusion center gathers and uses the \textit{examples}, which in this case are the agent observations. For each such example, the data tuple elements are the \textit{attributes} of the example.

\begin{definition}(Robust Concept)
    Let $\mathbf{y_i}$ be the actual data and $\tilde{\mathbf{y}}$ be %slightly off 
    the modified data. Let $r\text{ }(>0)$ be a robustness parameter. Consider $F(\mathbf{y_i})$ and $F(\tilde{\mathbf{y}})$ be the concepts deduced from $\mathbf{y_i}$ and $\tilde{\mathbf{y}}$, respectively. The concept is said to be $r$-robust if for $r>0, \exists \; \epsilon >0$ such that
    \[
        ||F(\mathbf{y_i})-F(\tilde{\mathbf{y}})||< \epsilon \;\; \text{whenever}\;\; 0<||\mathbf{y_i}-\tilde{\mathbf{y}}||<r.
    \]
\end{definition}

A concept is said to be $r$ robust if it can be accurately inferred even when the input example is perturbed within a bounded range defined by \textit{r}. That is, small deviations, such as minor reductions in the number of attributes or slight modifications to attribute values, do not substantially affect the correctness of the inferred concept. In the context of a MAS, the agent observations may be subject to local perturbations (e.g., due to privacy-preserving transformations), yet the fusion center must still be able to reliably deduce the parameter.

Robustness is a property inherent of the concept from an estimation perspective. The robustness parameter governs the tolerance to input perturbations and influences the success or failure of concept inference. Thus, it is typically specified a priori; we consider the value of $r$ to be $l$, the length of the square grid cell. 

\subsection{Compression Methodology: Random Projection}\label{Compression_Method}
Basic Random Projection (BRP) \cite{Random_Proj1}  is a dimensionality reduction mechanism where an \textit{n-}tuple data vector is multiplied with an $n\times m$ uniform random orthonormal matrix $A$ to get an \textit{m-}tuple. A fixed matrix is used for the projection of all data tuples.
Using this mechanism, $T(\mathbf{y_i})$ introduced in Equation \ref{eqn:T(y)} can be depicted mathematically as 
\begin{equation}
    T(\mathbf{y_i})=A^T\mathbf{y_i}.
\end{equation} 

One important property of BRP is that it preserves pairwise distances when the matrix $A$ is a uniform random orthonormal matrix. Using the lemma of Johnson and Lindenstrauss\cite{Frankl_Maehara}, Vempala\cite{Random_Proj1} has shown that the distance between two pairs of projections is bounded, with high probability, by a factor of the distance between the original data points. 
We have reformulated the same as Lemma~\ref{RP_lemma1} 
to highlight the similarity between distance preservation and differential privacy guarantee.
\begin{lemma}\label{RP_lemma1}
     Let $\mathscr{S} \subset \mathbb{R}^n$ be a set of data points with $|\mathscr{S}|=N$, where each point corresponds to the data from an agent. 
     For any $\gamma \in(0, 0.405)$, consider an uniform random projection to an $m$-dimensional subspace, where $m$ is suitably chosen, the following holds: \\
    For every pair $\mathbf{y_i}, \mathbf{y_j} \in \mathscr{S}$, 
    \begin{equation}\label{eqn:distance_preserving}
    \begin{split}
        P(e^{-\gamma}||\mathbf{y_i}-\mathbf{y_j}||^2 &\leq ||T(\mathbf{y_i})-T(\mathbf{y_j})||^2 \leq e^{\gamma}||\mathbf{y_i}-\mathbf{y_j}||^2)\\
         &\geq 1 - 2\sqrt{m}e^{-(m-1)(\frac{\sinh^2{\gamma}}{4}-\frac{\sinh^3{\gamma}}{6})}.
    \end{split}
    \end{equation} where $T(\mathbf{y_i})$ and $T(\mathbf{y_j})$ are the projections of $\mathbf{y_i}$ and $\mathbf{y_j}$ respectively.
\end{lemma}

Following a result due to Vempala \cite{Random_Proj1}, if we choose the projection dimension \textit{m} such that
\[
m \geq \frac{9\ln{N}}{\sinh^2{\gamma} - \frac{2}{3}\sinh^3{\gamma}}+1, 
\]
and substitute in the right-hand side of inequality \ref{eqn:distance_preserving}, we have 
\begin{equation}
    1 - 2\sqrt{m}e^{-(m-1)(\frac{\sinh^2{\gamma}}{4}-\frac{\sinh^3{\gamma}}{6})} \geq \frac{1}{2}.
\end{equation}
This result implies that the pairwise distances between projected data points are preserved within a multiplicative factor of $e^{\pm\gamma}$ with probability of at least $\frac{1}{2}$.

The distance preserving property of random projection ensures that the data vectors close to each other do not get mapped to vectors far apart in the lower dimensional space. We assume that the metric used for the calculation of pairwise distances is the Euclidean metric. This property is critical to maintaining the utility of sanitized data. However, the generation of a orthonormal matrix required for such projections incurs significant computational cost, typically on the order of $O(n\times m^2)$ \cite{Ortho_matrix}. Given the limited computational capacity and energy constraints of agents in a MAS, it may not always be feasible for each agent to independently generate and apply a random orthonormal projection for data sanitization. To address this limitation, we next introduce a modification of the BRP technique that retains a similar distance-preserving property, but is considerably more efficient and practical for deployment in resource-constrained environments.

\subsection{Neuron-friendly or Neuronal random projection}
It has been established that, for the purpose of dimensionality reduction, it is sufficient to use random matrices whose entries are independently drawn from a bounded distribution~\cite{Random_Proj1}. Such matrices, though not orthonormal, are capable of approximately preserving pairwise distances. Lemma~\ref{neuronal_lemma1} formalizes the distance preserving property. More specifically, like the previous lemma, it shows that for any pair of input vectors, their squared Euclidean distance is preserved up to a multiplicative factor $e^{\pm\gamma}$. 

\begin{lemma}\label{neuronal_lemma1}
Let $\mathbf{y_i}, \mathbf{y_j} \in \mathbb{R}^n$. Let $T(\mathbf{y_i})$, $T(\mathbf{y_j})$ be the projections of $\mathbf{y_i}$ and $\mathbf{y_j}$ to $\mathbb{R}^m$ via a random matrix $A$ with independent entries from a bounded distribution. Then 
\begin{equation}
    \begin{split}
    P(e^{-\gamma}||\mathbf{y_i}-\mathbf{y_j}||^2 &\leq ||T(\mathbf{y_i})-T(\mathbf{y_j})||^2 \leq e^{\gamma}||\mathbf{y_i}-\mathbf{y_j}||^2)\\
    &\geq 1 - 2e^{-(\sinh^2{\gamma}-\sinh^3{\gamma})\frac{m}{4}}.
    \end{split}
\end{equation}
\end{lemma}

For $\gamma \in (0, 0.405)$, the probability of pairwise distance preservation is higher in Lemma \ref{neuronal_lemma1} as compared to the preceeding lemma.

To address the computational limitations associated with generating orthonormal matrices in standard random projection, we propose an alternative approach, neuronal random projection (NRP). The sanitization process based on NRP is outlined in Algorithm \ref{alg:Sanitization_mechanism_neuronalRP}.
The nomenclature NRP draws inspiration from human anatomy, wherein neurons (analogous to agents) communicate information to the brain (analogous to the fusion center) to deduce and disseminate complex concepts. In this analogy, the generation of an orthonormal matrix by each neuron would represent a computationally intensive task, much like the high resource demands imposed on agents. In contrast, the NRP method is computationally lightweight, thereby offering a neuron-friendly alternative that is both privacy-preserving and efficient.

In NRP, each agent locally generates a random projection matrix with independent entries drawn from a bounded distribution. The matrix is further constrained to satisfy the norm bound derived in Theorem~\ref{Lemma:Norm_bound}, thereby ensuring compliance with the desired utility-privacy trade-off. Unlike the BRP method, where an agent may use the same compression matrix across multiple data tuples, the NRP approach involves generating a fresh projection matrix for each sanitization instance. This per-instance randomness enhances the unpredictability of the transformation, thereby contributing to a stronger privacy guarantee.

\begin{algorithm}[H]
\caption{Neuronal random projection (NRP)}\label{alg:Sanitization_mechanism_neuronalRP}
\begin{algorithmic}[1]
\Require %\textbf{Input:}
     \textit{Data tuple $\mathbf{y_i}$ of length n, %Number of agents S, 
     Reduced dimension value m}
\Ensure %\textbf{Output:}
    \textit{Data tuple $T(\mathbf{y_i})$ of dimension $m$}
\State $A[i][j] \gets 0$ \Comment{Initializing the compression matrix to zero}
\State $D[i] \gets 0$ \Comment{For random independent vectors forming R}
\State \textbf{Procedure:}{}
        \For{{$i \gets $1} to  $m$}
             \State $D[i] \gets$ Random independent vector of length $n$ from bounded distribution
        \EndFor
\State $A \gets $ Matrix whose columns are the vectors $D[i]$
\State{Compute $A^T\mathbf{y_i}$}
\State{Set $T(\mathbf{y_i})=A^T\mathbf{y_i}$}

\end{algorithmic}
\end{algorithm}

Based on Lemma \ref{RP_lemma1} and Lemma \ref{neuronal_lemma1}, in Theorem \ref{thm:nrp}, we also establish that the NRP method provides the same privacy guarantee as that of conventional random projection. 
\begin{theorem}\label{thm:nrp} The probabilistic guarantee for Neuronal random projection to satisfy the distance preservation property is as good as standard random projection.
\end{theorem}
\begin{proof}
    Let the probability of satisfying the distance preservation property be equal when $m$ is $m_1$ for basic random projection and $m_2$ for neuronal random projection.\\
    Solving the following equation:
    \begin{equation*}
        1 - 2\sqrt{m_1}e^{-(m_1-1)(\frac{\sinh^2{\gamma}}{4}-\frac{\sinh^3{\gamma}}{6})}=1 - 2e^{-(\sinh^2{\gamma}-\sinh^3{\gamma})\frac{m_2}{4}},
    \end{equation*}
    We have 
    \begin{equation*}
        m_2=\dfrac{(m_1-1)(\sinh^2{\gamma}-\frac{2}{3}\sinh^3{\gamma})-2\ln{m_1}}{\sinh^2{\gamma}-\sinh^3{\gamma}}.
    \end{equation*}
    Since $\gamma \in (0, 0.4054651081)$, we have $m_2 < 1.243m_1$. Since $m_2$ is the reduced dimension, it has to be an integral value. This implies, $m_2 \leq m_1$.
\end{proof}

\begin{figure}
    \centering
    \includegraphics[width=\linewidth]{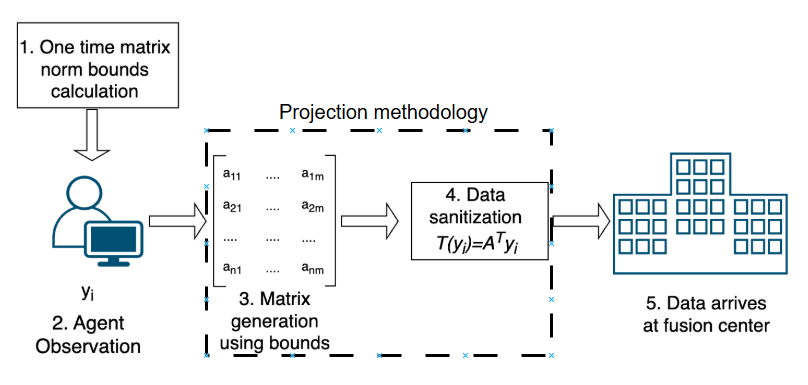}
    \caption{Outline of the sanitization process}
    \label{fig:Methodbrief}
\end{figure}

 Finally, we outline the general workflow for the preservation of inference privacy, as illustrated in Figure~\ref{fig:Methodbrief}. Each agent computes a bound on the norm of the projection matrix. This bound is derived on the basis of the maximum possible norm of its observation tuples and the size of the grid cell. In every subsequent sanitization instance, the agent generates a new projection matrix constrained by this norm bound. Each observation is then sanitized via matrix multiplication, thereby achieving dimensionality reduction in accordance with the utility-privacy trade-off. The use of a new matrix for each observation enhances randomness and prevents linkability between multiple data points.

\section{Experiment and results} \label{sec:expt}
This section first presents a detailed comparison of the computational complexities of our proposed method with three privacy-preserving approaches: PCA, ASUP and BRP. Once the efficiency of our method in case of large number of records and agents is established, we compare our algorithm experimentally with the other algorithms. 
\subsection{Time Complexity Comparison}
For this comparison, we assume $N$ to be the total number of data tuples for one agent, $n$ to be the original dimension of the data tuple, and $m$ to be the reduced dimension. 
\begin{enumerate}
    \item \textbf{Principal Component Analysis (PCA) \cite{New_PCA}}: The algorithm involves training each node independently on its local data before sharing the data with the fusion center. This process involves centering the data, computing the covariance matrix and the Eigen decomposition. These steps have a time complexity of $O(Nn)$, $O(Nn^2)$ and $O(n^3)$ \cite{QR_complexity}, respectively. This results in a preprocessing time complexity of $O(Nn^2+n^3)$. Subsequent projections of each new datapoint by matrix multiplication from $n$ to $m$ dimension have a time complexity of $O(nm)$.
    \item \textbf{Arbitrarily strong privacy with perfect utility sanitization algorithm (ASUP) \cite{ASUP}}: This algorithm requires choosing an $n\times n$ orthonormal basis (by QR decomposition or Gram-Schmidt orthonormalization \cite{QR_complexity}), followed by the identification of a vector not in the subspace spanned by the matrix, and the subsequent construction of a unitary matrix. These steps have time complexities of $O(n^3)$, $O(n^2)$, and $O(n^3)$, respectively. This is succeeded by the setting up of diagonal matrices that satisfy the norm constraints. This process has a linear time complexity. The final step of generating Gaussian noise, in which the unitary matrix ensures that the noise is rotated properly according to the desired Eigen structure of the covariance matrix has $O(n)$ complexity. All the steps lead to a cubic time complexity for ASUP, that is, $O(n^3)$.
    \item \textbf{Basic Random Projection (BRP) \cite{Random_Proj1}}: BRP involves the generation of an orthonormal matrix and the use of that matrix for the subsequent projection of all data tuples. The generation of a $n \times m$ orthonormal matrix has a time complexity of $O(nm^2)$. The projection of a $n$-tuple using the $n \times m$ takes $O(nm)$ time.
    \item \textbf{Proposed NRP with matrix norm bounds constraint}: Each tuple is projected using a fresh $n \times m$ matrix with entries drawn independently from a bounded distribution. The matrix is then scaled to satisfy the norm bounds. This process takes $O(nm)$ time as a $n \times m$ can be populated in $O(nm)$ time. The $n$-tuple is then projected using this matrix to an $m$-tuple in $O(nm)$ time.
\end{enumerate}

\begin{table}[h!]
\centering
\begin{tabular}{|c|cc|cc|}
\hline
 & \multicolumn{2}{c}{} & \multicolumn{2}{c|}{}\\
\multirow{4}{*}{Algorithm} & \multicolumn{4}{c|}{Time Complexity}  \\
 & \multicolumn{2}{c}{} & \multicolumn{2}{c|}{}\\
\cline{2-5}
 & \multicolumn{2}{c|}{} & \multicolumn{2}{c|}{}\\
 & \multicolumn{2}{c|}{Pre-processing} & \multicolumn{2}{c|}{Sanitization}\\
  & \multicolumn{2}{c|}{} & \multicolumn{2}{c|}{}\\
\hline
 & \multicolumn{2}{c|}{} & \multicolumn{2}{c|}{}\\
PCA & \multicolumn{2}{c|}{$O(Nn^2+n^3)$} & \multicolumn{2}{c|}{$O(nm)$}\\
 & \multicolumn{2}{c|}{} & \multicolumn{2}{c|}{}\\
\hline
 & \multicolumn{2}{c|}{} & \multicolumn{2}{c|}{}\\
ASUP &  \multicolumn{2}{c|}{--} & \multicolumn{2}{c|}{$O(n^3)$}\\
 & \multicolumn{2}{c|}{} & \multicolumn{2}{c|}{}\\
\hline
 & \multicolumn{2}{c|}{} & \multicolumn{2}{c|}{}\\
BRP & \multicolumn{2}{c|}{$O(nm^2)$} & \multicolumn{2}{c|}{$O(nm)$}\\
 & \multicolumn{2}{c|}{} & \multicolumn{2}{c|}{}\\
\hline
 & \multicolumn{2}{c|}{} & \multicolumn{2}{c|}{}\\
Proposed NRP & \multicolumn{2}{c|}{--} & \multicolumn{2}{c|}{$O(nm)$}\\
 & \multicolumn{2}{c|}{} & \multicolumn{2}{c|}{}\\
\hline
\end{tabular}\caption{Time complexity comparison of PCA, ASUP, BRP and our proposed method.}\label{tab:time_complexity}
\end{table}

The time complexities of the algorithms have been divided into pre-processing and sanitization complexities and tabulated in Table \ref{tab:time_complexity}. As evident, the time complexity of our algorithm is similar to that of PCA and BRP, but our method does not have an additional pre-processing step. This leads to better applicability of our proposed method to practical MAS models. We now present the results of the experimental comparison with ASUP, PCA and BRP, to corroborate our theoretical analysis. 
\subsection{Adversary Model}
We consider a passive, honest-but-curious adversary operating at the fusion center. The adversary follows the communication protocol and receives the sanitized data tuples $T(\mathbf{y_i})$ from each agent $i$. It then attempts to reconstruct the original observation vector $\mathbf{y_i}$ and further infer secondary private attributes from them.
\subsubsection*{(a) Adversary Capabilities:}
\begin{itemize}
    \item \textbf{Knowledge of the privacy mechanism:} The adversary is aware that Neuronal Random Projection (NRP) is used, including the distribution from which the matrices are generated.
    \item \textbf{Access to sanitized data:} The adversary has full access to all sanitized tuples $\{T(\mathbf{y_1}), T(\mathbf{y_2}), \ldots, T(\mathbf{y_N})\}$.
    \item \textbf{No access to raw data or compression matrix:} The adversary does not know $\mathbf{y_i}$ or the specific compression matrix $A_i$ used by agent $i$. Since each matrix is generated afresh for every sanitization step, reconstructing $A_i$ is computationally infeasible.
    \item \textbf{Inference capability:} The adversary employs reconstruction based data backtracking such as matrix inversion to estimate the original vector $\mathbf{y_i}$, particularly the private components.
\end{itemize}
\subsubsection*{(b) Adversary Objectives:}
\begin{itemize}
    \item \textbf{Reconstruction:} Estimate the raw data vector $\mathbf{y_i}$ from its compressed version $T(\mathbf{y_i})$. Estimation of the raw vector would help in secondary data inference.
\end{itemize}
This adversary model captures realistic insider threat scenarios in applications like smart healthcare, where the fusion center is protocol compliant but privacy intrusive. Our evaluation focuses on measuring the effectiveness of the proposed sanitization mechanism in limiting the adversary's ability to infer sensitive information. Our proposed NRP mechanism does so, by introducing per instance randomization and incorporating matrix norm bounds.

\subsection{Datasets} We experimented with the following datasets:
\begin{itemize}
    \item Synthetic data: The distribution function for the generation of the synthetic data is kept similar to that used by Wang et al.~\cite{ASUP}. The following assumptions are made to construct a sufficiently rich dataset for evaluation:
    \begin{itemize}
    \item Number of agents vary between 50 to 600 agents.
    \item Each agent makes 50 observations.
    \item The system parameter tuple $\mathbf{x}$ is 50 in length. 
    \item Each observation tuple has 12 private parameters and the rest are public.
    \item The entries in the observation model matrix are drawn independently from the uniform distribution \text{Unif}(-0.5, 0.5).

    \end{itemize}
    
    \item Real-life hospital data set: This data was collected from patients admitted over a period of two years\cite{Hospital_data}. The data set has multiple parameters including age, sex, platelet count, glucose level, and alcoholism. After the removal of few non-numeric fields and the conversion of binary non-numeric fields to binary numeric fields, the dataset is of length 50. We have considered identifying parameters such as age, gender, and duration of stay as private, while other health records are considered public parameters. 
\end{itemize}

\subsection{Algorithms used for comparison}
We compare our algorithm with the ASUP algorithm\cite{ASUP}, the PCA algorithm \cite{New_PCA}, and BRP\cite{Random_Proj1}. As the entire data set is available for evaluation, centralized methods such as PCA and ASUP can be applied for benchmarking purposes. Additionally, since BRP serves as the foundational approach for our NRP mechanism, we include it in our comparisons.

\subsection{Performance metrics}
The adversary is assumed to have complete knowledge of the privacy mechanism used by the agent, they apply a customized backtracking algorithm for the reconstruction process. In this study, we use a matrix multiplication-based reconstruction algorithm. In particular, as our NRP mechanism incorporates explicit constraints to ensure utility retention, the evaluation focuses primarily on measuring privacy. Specifically, our objective is to determine which mechanism offers the strongest resistance to inference attacks. For this purpose, we employ a set of performance metrics that compare the original data with the reconstructed data. These metrics are adapted from those introduced in \cite{SOCA} to suit our context.

\subsubsection{Breach count}
Breach count is the average number of data points guessed correctly by the adversary. We say that the adversary can breach the privacy of the user if he is able to reconstruct and place the datapoint in a neighbourhood of the original one.  
Let \textit{I} be an indicator variable that is set to 1 if the attacker correctly places the data tuple in a neighbourhood of the original tuple. Otherwise, the variable is set to 0. For the reconstructed dataset $\mathbf{A}^{'}=\{a_1',a_2', \ldots, a_n'\}$ and an actual dataset $\mathbf{A}=\{a_1,a_2, \ldots, a_n\}$ of size \textit{n}, the breach count is defined as:
\begin{equation*}
    \text{breach count} = \frac{\sum_{i=1}^{n} I_{(a'_{i} \in N(a_{i}))}}{n},
\end{equation*}where $N(a_{i})$ is the neighbourhood of $a_i$. A point $a_i'$ is said to belong to the neighbourhood $N(a_i)$ if the distance between the two points is atmost equal to the neighbourhood radius. The neighbourhood radius is set according to the threshold prespecified. Higher value of breach count implies lower levels of privacy, as the agent is able to correctly reconstruct the datapoints.
\subsubsection{Displacement} Displacement is defined as the average distance between the actual and the reconstructed data point. For a given reconstructed dataset $\mathbf{A'}$ and an actual dataset $\mathbf{A}$, the displacement is defined as 
\begin{equation*}
    \text{displacement} = \frac{\sum_{i=1}^{n} d(a_{i} , a_{i}^{'})}{n},
\end{equation*}
where $a' \in A'$ and $a \in A$. The displacement between two points is measured in terms of the Euclidean distance (i.e. the $l^{2}$ norm $\vert \vert.\vert \vert_{2}$). 

Higher the displacement between the actual and the reconstructed datapoints, higher is the level of privacy achieved as the agent is not able to correctly locate the cell that the actual datapoint belongs to. 
\subsubsection{Resemblance}\label{def:Resemblance}
This metric gives a measure of the nearest neighbours that are common to both the actual and reconstructed datapoints. More formally, let $S=\{\mathbf{s_1}, \ldots, \mathbf{s_k}\}$ be the set of \textit{k} nearest neighbors corresponding to the actual data point. Similarly, let $S^{'}=\{\mathbf{s'_1}, \ldots, \mathbf{s'_k}\}$ be the neighbors corresponding to the reconstructed data point. Then resemblance is defined as
\begin{equation*}
    \frac{\lvert S \cap S^{'} \rvert}{\lvert S \rvert}.
\end{equation*}

Lower resemblance implies higher levels of achieved privacy as the number of nearest neighbors that are common to both the actual and reconstructed data points is less in number. This would imply that an adversary is not able to correctly profile the agent using its neighbors.

\subsection{Numerical results}
To evaluate the effectiveness of incorporating matrix norm bounds in standard NRP, we have conducted an ablation study. This will highlight the impact of constraining the perturbation via matrix norms on privacy leakage. In this ablation, we systematically varied both the number of agents and the parameter $\epsilon$ used to derive the norm bounds, ensuring that the effect of incorporation of the matrix norm bound is robust across various configurations. Across all tested settings, privacy leakage was significantly mitigated in our proposed method relative to the standard NRP approach. A sample of the output is shown in Table \ref{tab:ablation} with 200 agents and $\epsilon$ at 0.5. As evident, the values obtained for our method is atleast 3 times better than that of standard NRP. This highlights the efficiency achieved by the incorporation of the matrix bounds into NRP.
\begin{table}[h!]
    \centering
    \begin{tabular}{|c|c|c|c|}
         \hline
  Dataset     & Metric & Standard NRP & Proposed NRP  \\
  \hline
    & & & \\
  Hospital Dataset & Breach Count  & 0.685 & 0.08 \\
    & & & \\
            & Displacement   & 1.805 & 19.789\\
    & & & \\
            & Resemblance & 0.4317  & 0.1767\\
    & & & \\
  \hline
    & & & \\
  Synthetic Dataset & Breach Count  & 0.66 & 0.11 \\
    & & & \\
            & Displacement   & 35.912 & 258.289\\
    & & & \\
            & Resemblance & 0.0219  & 0.0088\\
    & & & \\
  \hline
    \end{tabular}
    \caption{Comparison between standard NRP and our proposed NRP}
    \label{tab:ablation}
\end{table}

We compare our neuronal random projection mechanism with the three other mechanisms while varying the number of agents from 50 to 600. The value of $\epsilon$ has been taken to be 0.5. For the dimension reduction based methods, we have taken the reduced dimension $m$ to be 20. Other combinations of $\epsilon \text{ and } m$ yielded similar results. The numerical results are obtained by averaging the metric values over 100 executions of the algorithms. For reconstruction, we generate a random inverse matrix without prior knowledge of the compression matrix.

\begin{figure*}[h!]
\centering
\subfloat[]{\includegraphics[width=0.45\linewidth]{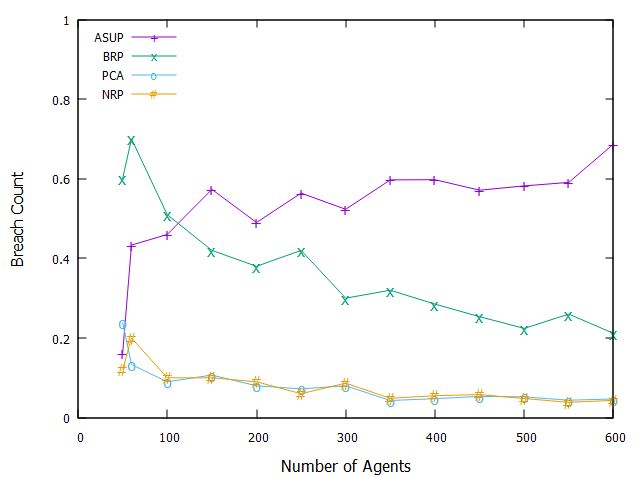}%
\label{fig1_first_case}}
\hfil
\subfloat[]{\includegraphics[width=0.45\linewidth]{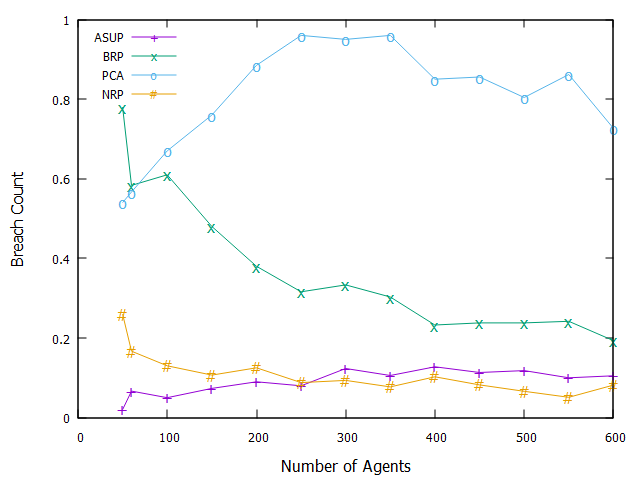}%
\label{fig1_second_case}}
\caption{Breach Count Comparison for (a) Real life data (b) Synthetic data}
\label{fig:breach}
\end{figure*}

\begin{figure*}[h!]
\centering
\subfloat[]{\includegraphics[width=0.45\linewidth]{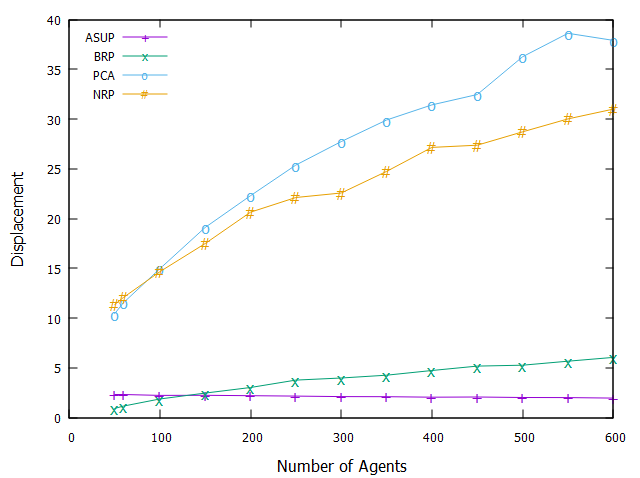}%
\label{fig2_first_case}}
\hfil
\subfloat[]{\includegraphics[width=0.45\linewidth]{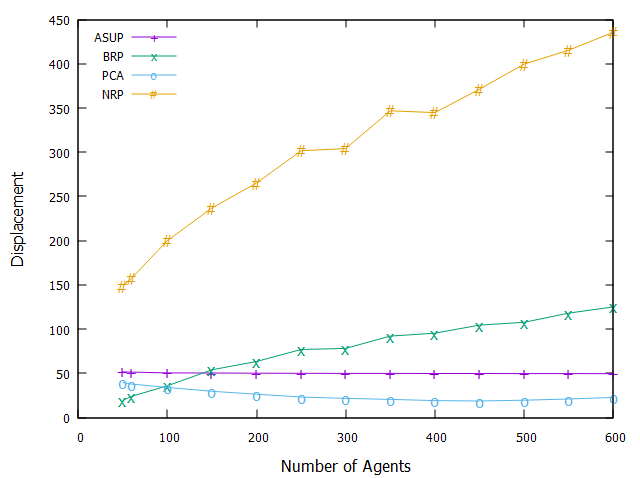}%
\label{fig2_second_case}}
\caption{Displacement Comparison for (a) Real life data (b) Synthetic data}
\label{fig:displacement}
\end{figure*}

\begin{figure*}[h!]
\centering
\subfloat[]{\includegraphics[width=0.45\linewidth]{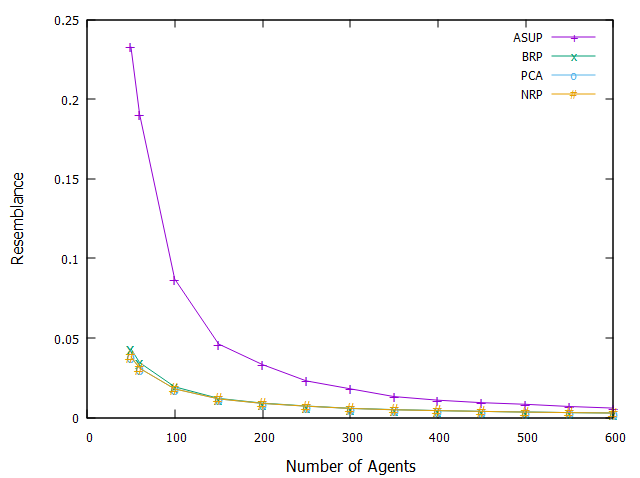}%
\label{fig3_first_case}}
\hfil
\subfloat[]{\includegraphics[width=0.45\linewidth]{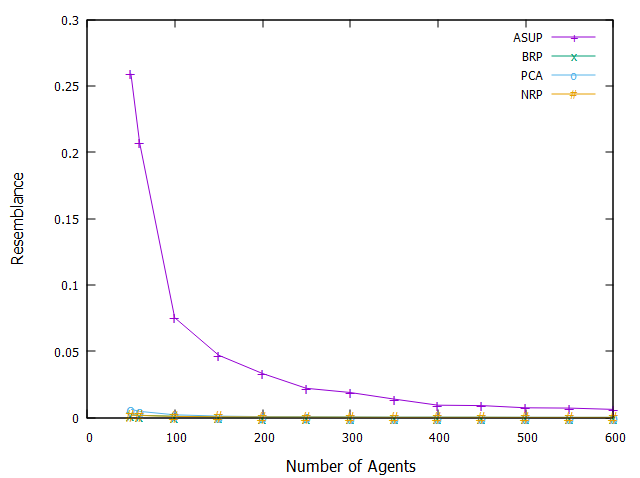}%
\label{fig3_second_case}}
\caption{Resemblance Comparison for (a) Real life data (b) Synthetic data}
\label{fig:resemblance}
\end{figure*}

Results for the real life dataset and the synthetic dataset are plotted in Figures \ref{fig:breach}-\ref{fig:resemblance} respectively. We next analyze the numerical results.

\subsubsection{Breach Count Analysis}
We have taken a threshold value of 20\% as the radius of the neighborhood. We have also experimented with other threshold values. As seen in Figure \ref{fig1_first_case}, for real-life data, our mechanism performs almost four times better than BRP, six times better than ASUP, and is almost similar to PCA. The average breach count for NRP for real-life data is 0.0805, while it is 0.0837 for PCA, 0.3757 for BRP, and 0.5253 for ASUP. These values translate to approximate data leakages of 8\% for both NRP and PCA, and about 35\% and 50\% for BRP and ASUP, respectively.

For the synthetic data shown in Figure \ref{fig1_second_case}, NRP again shows superior performance, approximately three times better than BRP, seven times better than PCA, and comparable to ASUP. The average breach count for NRP in this case is 0.11, compared to 0.3791 for BRP, 0.7992 for PCA, and 0.0902 for ASUP. In the case of synthetic data, ASUP performs because it is designed for randomly generated data points and does not account for correlations among data elements. In contrast, PCA retains components associated with dominant eigenvectors and thus performs well when the data set exhibits internal correlations, such as in real-life data. However, PCA's effectiveness diminishes for uncorrelated synthetic data. Our NRP mechanism, by incorporating both utility preservation and noise-based privacy guarantees, adapts well to both types of data.

\subsubsection{Displacement Analysis}
For the real-life dataset, ASUP exhibited the lowest displacement (Figure~\ref{fig2_first_case}), with an average value of 2.14. This indicates that ASUP achieves the least privacy among the compared mechanisms. On the contrary, NRP shows high displacement values (22.29), comparable to PCA (25.99), suggesting a significantly higher level of privacy preservation. BRP is between with a moderate displacement value of 3.72. 

In the case of the synthetic dataset (Figure~\ref{fig2_second_case}), PCA achieved the lowest displacement (25.78), again implying the least privacy protection. ASUP and BRP show slightly higher displacement values of 50.22 and 76.53, respectively. NRP records the highest displacement value (302.08), indicating strong privacy preservation. Compared to PCA, ASUP and BRP, NRP achieves an improvement of approximately 91\%, 83\%, and 75\%, respectively, in terms of privacy, as reflected by displacement.

\subsubsection{Resemblance Analysis}
We evaluate resemblance by comparing the sets of the 10 closest neighbors of the original and reconstructed datapoints. A lower resemblance value indicates better privacy, as it suggests a greater distortion of neighbor relationships during sanitization. 

For the hospital dataset, the average resemblance values for ASUP, PCA, BRP, and NRP are 0.05275, 0.01101, 0.01186, and 0.01102, respectively. This implies that PCA, BRP, and NRP reduce resemblance by approximately 79\%, 77\%, and 79\% compared to ASUP, respectively, indicating a significantly higher degree of neighbourhood distortion and, hence, better privacy preservation. In the case of the synthetic dataset, the resemblance values are 0.05507 for ASUP, 0.00132 for PCA, 0.00058 for BRP, and 0.00054 for NRP. These figures correspond to reductions of approximately 98\% for PCA, and over 99\% for both BRP and NRP relative to ASUP. Notably, NRP exhibits the lowest resemblance across both datasets, marginally outperforming PCA and BRP, thereby demonstrating its strong privacy-preserving capability.

In the noise-based method where the distribution parameters do not change over time, the sanitized data can only undergo a structured modification. Consequently, with a carefully designed reconstruction algorithm, the sanitized data points can be mapped so that the set of common neighbors of the reconstructed and original datapoints coincide. This results in higher resemblance values or lower privacy for the noise-based method ASUP.  
In contrast, dimension reduction methods such as PCA, BRP, and NRP inherently distort the neighborhood structure by projecting data into a lower-dimensional space, making exact neighbor preservation less likely. However, both PCA and BRP use a fixed projection matrix in all sanitization steps. This consistency increases the probability of some recurring neighborhood patterns, resulting in resemblance values that are slightly higher than those observed for NRP. The use of a fresh random projection matrix for each sanitization step in NRP introduces variability and unpredictability, significantly reducing the likelihood that a reconstructed data point shares a common set of neighbors with its original counterpart. This makes NRP particularly effective in minimizing resemblance and enhancing privacy.

\subsection{Discussion}
Among the mechanisms evaluated, ASUP, PCA and BRP demonstrate strengths that are largely dataset-dependent. ASUP, being a noise addition-based method, performs reasonably well for synthetic data. It is data set-specific and is good only for uncorrelated data.

PCA performs well on real-life datasets where the data exhibit internal correlations. By identifying and preserving the principal components that capture the most variance, PCA ensures high utility and moderate privacy. However, in the absence of meaningful correlations, as is the case with randomly generated synthetic data, it fails as its assumptions about data structure no longer hold.
 
BRP applies a fixed projection matrix to all datapoints, making it data-independent. Although this allows it to function equally well across both datasets, its privacy performance is suboptimal due to the rigidity of the projection matrix.

In contrast, our proposed NRP consistently outperforms the others in all three privacy metrics for both datasets.  It introduces randomness by generating a fresh projection matrix at each sanitization step, thereby making reconstruction significantly more difficult for an adversary. This approach ensures strong privacy guarantees while retaining utility, irrespective of the data's inherent structure.

Another observation is that compression-based mechanisms perform better in terms of reducing resemblance. This is because they project data into a lower-dimensional space, and without knowledge of the exact transformation matrix, it is computationally infeasible for an adversary to reconstruct data points that preserve original neighbourhoods.

\section{Conclusion and Future Work}\label{Future}
In this paper, we proposed a novel neuronal random projection (NRP) mechanism to preserve inference privacy. Through extensive evaluation on real-life and synthetic datasets, we demonstrated that NRP consistently outperforms existing mechanisms. Unlike basic random projection (BRP), which requires an orthonormal projection matrix, we established that it is sufficient to use random matrices with independently drawn entries from a bounded distribution. This relaxation not only improves privacy through randomization, but also significantly reduces computational complexity.

While the proposed NRP mechanism demonstrates strong performance, it is not without limitations. The effectiveness of NRP heavily relies on the assumption that the random projection matrices, satisfy the derived matrix norm bounds. If the input data is sparse or has extreme variability, repeated generation of such matrices will require rejection sampling or scaling, thereby introducing runtime overhead. While the randomized projection enhances privacy, it may also hinder tasks requiring data consistency across multiple time steps. Finally, NRP is designed for passive adversaries with no background knowledge or adaptivity. In future work, we plan to extend the framework to address multiple threat models, including active adversaries and adversaries with prior knowledge of the data distributions.

We also wish to devise a tunable sanitization mechanism with the provision of specifying the privacy level desired. We want the number of private and public parameters to be enforced agent-wise and not uniform for all agents. Additionally, we plan to extend our framework to dynamic environments, where agents may join or leave the system over time. 

\end{sloppypar}

\end{document}